%% file: ms.tex
\pdfoutput=1
\documentclass[11pt]{article}

\usepackage[]{acl}

\usepackage{times}
\usepackage{latexsym}
\usepackage{tabularx}
\usepackage{booktabs}
\usepackage{multirow}
\usepackage{xspace}
\usepackage{breqn}
\usepackage{amsmath}

\usepackage{cleveref}
\crefformat{section}{§#2#1#3}

\usepackage[T1]{fontenc}
\usepackage[utf8]{inputenc}

\usepackage{microtype}

\newcommand{\storc}{\textsc{S2ORC}\xspace}
\newcommand{\scibert}{\textsc{SciBERT}\xspace}
\newcommand{\scidocs}{\textsc{SciDocs}\xspace}
\newcommand{\scincl}{\textsc{SciNCL}\xspace}
\newcommand{\aspire}{\textsc{Aspire}\xspace}
\newcommand{\specter}{\textsc{Specter}\xspace}
\newcommand{\map}{\textsc{MAP}\xspace}
\newcommand{\ndcg}{\textsc{nDCG}\xspace}
\newcommand{\relish}{\textsc{relish}\xspace}
\newcommand{\treccovid}{\textsc{treccovid}\xspace}
\newcommand{\csfcube}{\textsc{CSFCube}\xspace}
\newcommand{\mdcr}{\textsc{MDCR}\xspace}

\title{
Large-scale Evaluation of Transformer-based Article Encoders\\
on the Task of Citation Recommendation}

\author{Zoran Medi\'{c} \and Jan \v{S}najder\\
Text Analysis and Knowledge Engineering Lab\\
Faculty of Electrical Engineering and Computing, University of Zagreb\\
Unska 3, 10000 Zagreb, Croatia \\
\tt \{zoran.medic,jan.snajder\}@fer.hr
}

\begin{document}
\maketitle

\input{0-Abstract}
\input{1-Introduction}
\input{2-ModelsBenchmarks}
\input{3-LargeScaleEvaluation}
\input{4-MultiDomainBenchmark}
\input{5-Conclusion}

\section*{Acknowledgments}

This research was supported by the grant AIDWAS KK.01.2.1.02.0285. The first author has been supported by a grant from the Croatian Science Foundation (HRZZ-DOK-2018-09).

\bibliography{anthology,custom}
\bibliographystyle{acl_natbib}

\input{6-Appendix}

\end{document}

%% file: 0-Abstract.tex
\begin{abstract}
Recently introduced transformer-based article encoders (TAEs) designed to produce similar vector representations for mutually related scientific articles have demonstrated strong performance on benchmark datasets for scientific article recommendation.
However, the existing benchmark datasets are predominantly focused on single domains and, in some cases, contain easy negatives in small candidate pools. 
Evaluating representations on such benchmarks might obscure the realistic performance of TAEs in setups with thousands of articles in candidate pools.
In this work, we evaluate TAEs on large benchmarks with more challenging candidate pools.
We compare the performance of TAEs with a lexical retrieval baseline model BM25 on the task of citation recommendation, where the model produces a list of recommendations for citing in a given input article.
We find out that BM25 is still very competitive with the state-of-the-art neural retrievers, a finding which is surprising given the strong performance of TAEs on small benchmarks.
As a remedy for the limitations of the existing benchmarks, we propose a new benchmark dataset for evaluating scientific article representations: \textbf{M}ulti-\textbf{D}omain \textbf{C}itation \textbf{R}ecommendation dataset (\textbf{\mdcr}), which covers different scientific fields and contains challenging candidate pools.
\end{abstract}

%% file: 1-Introduction.tex
\section{Introduction}
\label{sec:intro}

The introduction of large pre-trained language models (LMs) \citep{devlin-etal-2019-bert,radford2019language,lewis-etal-2020-bart,2020t5} based on the transformer architecture \citep{NIPS2017_3f5ee243} has improved performance on numerous NLP tasks.
The adaptation of LMs to scientific corpora \citep{beltagy-etal-2019-scibert,luu-etal-2021-explaining,gupta2022matscibert,lee2020biobert} laid the foundation for applying transformer-based LMs to various scholarly document processing (SDP) tasks, such as named-entity recognition \citep{naseem2020biomedical}, article summarization \citep{cai2022covidsum}, scientific fact-checking \citep{wadden-etal-2020-fact}, describing relationships between articles \cite{luu-etal-2021-explaining}, and citation recommendation (CR) \citep{nogueira2020evaluating,gu2022local}, among others.

While some of the SDP tasks rely on word- or sentence-level representations, others, such as CR and article summarization, require document-level representations.
To obtain such representations, recent work has proposed various transformer-based article encoders (TAEs), i.e., LMs that are finetuned using citation or co-citation information as a training signal, such as \specter \citep{cohan-etal-2020-specter}, \aspire \citep{mysore2021aspire}, and SciNCL \citep{ostendorff2022neighborhood}.
Representations obtained with these models can then be used in various downstream recommendation tasks where a user searches for articles that are in some way relevant to a given query article.

To date, article representations obtained with TAEs have been evaluated against recommendation benchmarks such as \scidocs \citep{cohan-etal-2020-specter}, \relish, \citep{brown2019large} or \treccovid \citep{treccovid}.
While \scidocs focuses mainly on the field of computer science, \relish and \treccovid cover articles from the biomedical field.
These benchmarks contain a set of query articles, where each query is paired with a candidate pool consisting of both relevant and irrelevant articles for that query.
The difference between the benchmarks, apart from the domains they cover, is how the candidate pools are constructed: 
\relish and \treccovid contain expert-annotated relevance labels for each candidate in a pool, while \scidocs uses random sampling of negative candidates for pool construction. 
However, they all contain relatively small candidate pools (e.g., 25 in \scidocs).
Such small and, in some cases, randomly sampled candidate pools do not resemble typical use-case scenarios in which query articles are compared to millions of candidate articles from public databases.
Thus, evaluating TAEs on such benchmarks can lead to an overly optimistic performance estimation as the candidate pool does not constitute a representative sample of the population of candidate articles in realistic use cases.

In this work, we turn to a more realistic evaluation of TAEs and evaluate them on large (\(\geq\)200k) candidate pools and across different scientific fields.
Although emulating such a realistic setup has so far been avoided due to the prohibitive computational cost of nearest neighbor search on millions of embeddings, research on GPU-based nearest neighbor (NN) search \citep{johnson2019billion} has given rise to efficient techniques that enable embedding-based search in large-scale setups.
To make use of fast NN search, we focus on the bi-encoder models \citep{lin2021pretrained}, that can be easily coupled with fast GPU-based NN search. We evaluate TAEs on the task of CR, in which a model outputs a list of articles as recommendations for citing in a given article.
Alongside TAEs, we evaluate the traditional lexical retrieval model BM25 \citep{robertson1994some}, which, in spite of its simplicity, still stands as a hard-to-beat baseline in many retrieval tasks.
Our evaluation shows that BM25 performs on par with TAEs in this setup, especially as candidate pools grow.

Building on the results of our large-scale evaluation of TAEs, we then construct a new benchmark dataset for evaluating scientific article representations on the task of CR.
Our Multi-Domain CR-based benchmark dataset (MDCR), albeit comparable in size to previous benchmarks, spans different scientific fields and consists of challenging candidate pools.
More precisely, candidate pools in \mdcr contain different candidate types, ranging from those obtained from the large-scale evaluation of state-of-the-art TAEs to candidates from the citation graph neighborhood.

To summarize, the contribution of our work is twofold: (1) we conduct a large-scale evaluation of state-of-the-art
TAEs on pools of varying sizes, and (2) present a new and challenging multi-domain benchmark dataset for evaluating scientific article representations that contains challenging candidates identified in the large-scale evaluation.\footnote{Our code, the data splits and the new benchmark data are publicly available at the following link: \url{https://github.com/zoranmedic/mdcr}.}

The rest of the paper is organized as follows.
In Section~\ref{sec:data_models}, we describe the models we evaluate and give an overview of the existing benchmarks for scientific article recommendation.
Section~\ref{sec:eval} presents the results of a large-scale evaluation of TAEs and BM25 in two evaluation setups.
In Section~\ref{sec:md-bench} we describe the construction of a new and more challenging multi-domain benchmark and present the initial results for the models we considered.
Section~\ref{sec:concl} concludes the paper and proposes future work.

%% file: 2-ModelsBenchmarks.tex
\section{Models and Benchmarks}
\label{sec:data_models}

\subsection{Transformer-based Article Encoders}
As a baseline TAE, we consider \textbf{\scibert} \citep{beltagy-etal-2019-scibert}, a variant of BERT \citep{devlin-etal-2019-bert}, trained on a corpus of scientific articles with masked language modeling objective.
Next, we include \textbf{\specter} \citep{cohan-etal-2020-specter}, a \scibert-based TAE trained with a contrastive learning objective that minimizes the L2 distance between embeddings of citing-cited article pairs.
Further, we consider \textbf{\scincl} \citep{ostendorff2022neighborhood}, another \scibert-based TAE that uses citation graph embeddings for a more informative selection of negative examples with the same contrastive learning objective as \specter.
Finally, we also evaluate \textbf{\aspire} \citep{mysore2021aspire}, a TAE that uses a co-citation signal to make sentence embeddings of co-cited articles similar.

Among these four TAEs, only \scibert is trained without any inter-article (i.e., citation or co-citation) training signal.
We thus consider it as a baseline to investigate how well LMs pre-trained on domain's corpora can be used in retrieval scenarios without any finetuning.
On the other hand, the rest of the TAEs differ both in the type of inter-article training signal used (co-citation for \aspire vs. citation for \specter and \scincl) and in the granularity of representation used for article matching (sentence embeddings for \aspire vs. document embeddings for \specter and \scincl). 
All the considered TAEs are \emph{bi-encoders} \citep{lin2021pretrained}, i.e., they produce dense representations of a single input article, which allows them to be easily employed in large-scale setups when coupled with fast nearest neighbor search methods. The alternative are the \emph{cross-encoders}, which take two concatenated articles as the input and output the relevance matching score. Although the cross-encoders often outperform bi-encoders, we do not consider them here as they are not compatible with nearest neighbor search methods and therefore not suitable for large-scale retrieval.
\footnote{We thus only consider \textsc{ts-Aspire} model in our work and leave out \textsc{ot-Aspire}, a variant that uses optimal transport over sentence embeddings, whose computational complexity prohibits its use in large-scale retrieval scenarios.}

All the considered TAEs produce scientific article representations using the article's title and abstract as input.
Since the title and abstract serve as a condensed overview of an article, it is clear that not all possible relationships between a pair of articles can be detected using such input only.
However, we consider the title and abstract a reliable proxy for otherwise complex and computationally expensive processing of the whole article's content.

\subsection{Existing Benchmarks}
Scientific article recommendation benchmarks that TAEs were evaluated on so far were designed for domain-specific retrieval evaluation across small-sized and, in some cases, randomly sampled candidate pools.
Each benchmark consists of a set of queries, where each query (title and abstract or a free-form text) is paired with a corresponding \textit{candidate pool}, i.e., a set of query-relevant (\textit{positive}) and query-irrelevant (\textit{negative candidates}) articles.
We review the most commonly used benchmarks below.
\begin{description}
\setlength\itemsep{0.8pt}
    \item \textbf{\scidocs} \citep{cohan-etal-2020-specter}: A collection of datasets for the evaluation of classification and retrieval tasks that use abstract-level article representations. In retrieval tasks, each query article is paired with a candidate pool of 5 positive and 25 randomly sampled negative candidates.
    \item \textbf{\relish} \citep{brown2019large}: A collection of query and candidate articles expert-annotated for relevance. Query articles are from the field of biomedicine, each paired with a set of 60 candidates.
    \item \textbf{\treccovid} \citep{treccovid}: A TREC-style benchmark consisting of various queries related to COVID-19. Each query is paired with around 300 candidate articles annotated for relevance by medical experts.
    \item \textbf{\csfcube} \citep{mysore2021csfcube}: An expert-annotated dataset of 50 computer science articles annotated at sentence-level for aspect-based relevance with candidate articles. The average candidate pool size is 125.
\end{description}

Three of these benchmarks (\relish, \treccovid, \csfcube) are single-domain by design, while \scidocs is constructed with queries from different scientific fields.
However, the majority of \scidocs queries (over 70\%) come from a single domain (computer science), making it a predominantly computer science-oriented benchmark.

Existing benchmarks also differ in how the candidate pools in each of them were constructed.
While \relish, \treccovid, and \csfcube contain expert-annotated candidate pools, meaning that field experts annotated the relevance of each candidate to the query, candidate pools for retrieval tasks in \scidocs are made of negative candidates randomly sampled from a set of articles that are not related to the query.
For example, in the case of the ``Cite'' task in \scidocs, each query article is paired with a pool of 5 articles cited in the query, and 25 negative candidates are randomly sampled from a held-out set of articles not cited in the query article.
An obvious advantage of random candidate pools over expert-annotated pools is that they are less expensive to construct.
However, a downside is that random candidate pools might contain many candidates that are entirely unrelated to the query and lead to overly optimistic performance estimates that are not representative of realistic large-scale retrieval scenarios.

%% file: 3-LargeScaleEvaluation.tex
\section{Large-Scale Evaluation}
\label{sec:eval}

We performed the large-scale evaluation in two setups: dataset- and field-level.
Dataset-level evaluation resembles a basic evaluation setup -- a random sampling of both queries and articles in the candidate pool.
The field-level evaluation focuses on specific scientific fields using queries and candidate pools comprised of articles from specific fields.

In both setups, we evaluated the chosen models on the task of global CR, in which a model is trained to produce a list of articles as recommendations for citing in a given query article.
Although CR is not the only task on which TAEs can be evaluated, it is arguably the most accessible among the article retrieving tasks. 
Whereas other tasks (e.g., user activity tasks) might require data that is typically not publicly available (e.g., search engine logs), CR datasets are easily obtained through parsing reference lists of publicly available articles.
Previous research on global CR has proposed many features that could be used to represent the input articles \citep{bhagavatula-etal-2018-content,ali2021global}.
However, in this work, we only use the article's title and abstract as input, as our focus is not on improving the state-of-the-art in global CR but rather on evaluating the TAE-produced article representations in a retrieval scenario.
For a detailed overview of the various tasks and methods in CR, we refer the reader to \citep{medic2020survey}.

For each TAE that we consider, the input was constructed by concatenating the input article's title and abstract (separated with a \texttt{[SEP]} token).
For \scibert, \specter, and \scincl, we used the final layer's \texttt{[CLS]} token embedding as input article's representation, while for \aspire we mean-pooled token embeddings across all layers for each sentence in the input.
We used HuggingFace's\footnote{\url{https://github.com/huggingface/transformers}} implementations of TAEs, while for BM25 we used Lucene's implementation, i.e., its Python toolkit \texttt{pyserini}.\footnote{\url{https://github.com/castorini/pyserini}}
For nearest neighbor search across article embeddings, we used Faiss \citep{johnson2019billion}.\footnote{\url{https://github.com/facebookresearch/faiss}}

We used the \storc dataset \citep{lo-etal-2020-s2orc} in all our experiments. 
\storc is a recently released large dataset of 81.1M scientific articles covering dozens of scientific fields.
Together with the metadata and article's title and abstract, the dataset contains citation links between the articles.
Therefore, we consider it appropriate for the large-scale evaluation, not just due to its size and coverage but recency as well.
We perform initial filtering of articles and remove all those with (1) empty publication year field, (2) empty title field, (3) abstract shorter than 30 characters, or (4) less than three citations in \storc.
This filtering leaves us with a \textit{prefiltered set} of around 16M articles that we use for both sampling of queries and candidate pool construction in both evaluation setups.

For both evaluation setups, we report the standard metrics used in prior work on scientific article recommendation: \map, \ndcg, and R@30.
We set \textit{k} in R@\textit{k} to 30, since on average there are 29 positives (cited articles) for each query in the query set.
All metrics range from 0 to 1, where higher is better.
Although defined differently, all the metrics yield higher values when relevant articles are positioned higher in the list of retrieved articles. 

\subsection{Dataset-level}
We start by describing the dataset-level setup in which we evaluated how TAEs perform when asked to provide recommendations over a large candidate pool for random queries from \storc.

First, we sampled a random set of 3800 query articles\footnote{In field-level setup, we sampled 200 queries for each of the 19 MAG fields. To keep the total number of queries the same over both setups, we sampled 3800 total queries in the dataset-level setup as well.} 
from the prefiltered set of articles.
We left out the query articles used in the training sets of \specter and \scincl.\footnote{At the time of writing, \aspire's training set was not publicly available, so we did not account for that overlap.}
Next, we sampled candidate pools of various sizes: 200k, 500k, 1M, and 2M.
Each candidate pool contained all the articles cited in the query articles, while the remaining candidates were randomly sampled from the prefiltered set.
To make the setup more realistic, we considered the publication years of both query and candidate articles: queries were sampled from the articles published in 2019, while candidate articles' year of publication was 2019 or earlier.
Year-based sampling ensures that no article published after the query article can be recommended for citing in that article.
Although such year-based sampling still allows for the articles published after the citing (later in 2019) to be included as candidates, it reduces such possibility compared to other benchmarks (e.g., \scidocs) that do not account for it.\footnote{Since \storc only provides publication years (and not the dates) for articles it contains, filtering can at most be year-based. Additionally, excluding from the candidate pools articles that were published in the same year as the citing would considerably reduce the pool size in some fields.}
For each candidate pool size, we repeat the pool sampling procedure three times and report the mean values of the metrics.

\begin{table*}[t]
\footnotesize
\setlength{\tabcolsep}{5.3pt}
\centering
\begin{tabular}{@{}lcccccccccccc@{}}
\toprule
Pool sizes $\rightarrow$ & \multicolumn{3}{c}{200k} & \multicolumn{3}{c}{500k} & \multicolumn{3}{c}{1M} & \multicolumn{3}{c}{2M} \\
\cmidrule(lr){2-4}\cmidrule(lr){5-7}\cmidrule(lr){8-10}\cmidrule(lr){11-13}
Models $\downarrow$ & \map & \ndcg & R@30 & \map & \ndcg & R@30 & \map & \ndcg & R@30 & \map & \ndcg & R@30 \\ \midrule
BM25 & 40.4 & 73.8 & 43.0 & 32.8 & 68.9 & 36.4 & \textbf{27.4} & \textbf{64.9} & 31.6 &\textbf{22.5} & \textbf{60.8} & \textbf{26.9} \\
\scibert & 5.5 & 40.3 & 5.6 & 4.6 & 38.5 & 3.9 & 4.1 & 37.3 & 3.0 & 3.8 & 36.5 & 2.2 \\
\specter & 37.4 & 72.0 & 40.9 & 29.5 & 66.5 & 33.9 & 24.1 & 62.1 & 28.7 & 19.2 & 57.8 & 23.8 \\
\scincl & \textbf{42.5} & \textbf{75.2} & \textbf{45.2} & \textbf{33.4} & \textbf{69.3} & \textbf{37.6} & 27.1 & 64.6 & \textbf{31.9} & 21.6 & 60.0 & 26.5 \\
\textsc{Aspire-BM} & 41.4 & 74.7 & 43.3 & 32.6 & 68.9 & 35.9 & 25.7 & 63.5 & 30.5 & 20.4 & 59.0 & 25.2 \\ \bottomrule
\end{tabular}
\caption{Results on different pool sizes in the dataset-level setup for BM25 and three considered TAEs. Values in \textbf{bold} indicate the best-performing model for a combination of pool size and metric.}
\label{tab:results-dataset-level}
\end{table*}

Dataset-level results are given in Table~\ref{tab:results-dataset-level}.
We retrieved the top 500 ranked candidates for each model and reported \map, \ndcg, and recall at 30 averaged over three runs for each pool size. 
For \aspire, we used its ``BioMed'' variant, i.e., the one trained on articles from the biomedicine field.\footnote{The other available \aspire model, trained on computer science articles, obtained worse results.}
We optimized BM25's parameters $b$ and $k_1$ on separate validation sets constructed in the same way as test sets.
A detailed description of the BM25 formula and the role of the two parameters is given in the Appendix \ref{sec:appendix}.

We observe that the best performing model on the pool sizes of 200k and 500k is \scincl, with \aspire and BM25 not far behind.
However, with larger pool sizes of 1M and 2M, BM25 performs better than TAEs for most metrics (except R@30 in the 1M pool, where \scincl outperforms BM25).
Given the slight difference in performance between BM25 and \scincl, our results demonstrate that traditional lexical retrieval is still very competitive in large-scale retrieval scenarios. 
These results are in line with those of \citep{reimers-gurevych-2021-curse}, who also compared the performance of sparse and dense retrieval models on varying pool sizes and found that the performance of the dense retrieval models decreases quicker for the increasing pool sizes compared to sparse methods.
Looking at differences between TAEs, the results show clear benefits of finetuning TAEs with inter-article training signal -- both \specter and \scincl outperform \scibert.

We also observed a significant drop in performance for all the evaluated TAEs compared to their performance on the ``Cite'' task in \scidocs (results on \scidocs are given in Appendix~\ref{sec:appendix}). 
For example, MAP for \scibert in the ``Cite'' task of \scidocs was 48.3 \citep{cohan-etal-2020-specter}, while in a large-scale setup, it ranges from 5.5 in the case of 200k pool size to 3.8 with a 2M pool size.
This difference supports our hypothesis that small-scale evaluation is not indicative of the performance of a model in a realistic, large-scale setup.
However, our large-scale evaluation results are consistent with some other findings from the evaluation on \scidocs, as reported in \citep{ostendorff2022neighborhood}: \scincl's careful sampling of negatives for the training set leads to a clear improvement in retrieval performance, with \scincl outperforming \specter for all candidate pool sizes.

\subsection{Field-level}
\label{subsec:field-level-eval}

In the field-level evaluation, we evaluate TAEs on a set of queries and candidate pools from specific scientific fields.
Such an evaluation setup resembles a more realistic and also more challenging large-scale retrieval scenario: in a real-world application, given a query article as input, a retrieval model is expected to detect the query article's field and narrow the candidate pool to articles from that field.

To determine the article's field, we used Microsoft Academic Graph (MAG) labels provided in \storc.
We sampled 200 query articles for each of the 19 distinct MAG fields from \storc.
As in the dataset-level setup, we used year-based splits and sample query articles published in 2019.
Next, for each scientific field, we constructed a candidate pool of size 100k that contains all the articles cited in the query articles alongside field-specific negative candidates.
To obtain field-specific negative candidates, we randomly sampled the remaining pool articles (up to 100k) from a set of field-cited articles, i.e., a set of articles cited in all \storc articles labeled with a specific MAG field.
For example, when sampling negative candidates for the Medicine field, we first filtered all articles labeled with Medicine in their \storc's MAG field.
We then went through all the articles that they cite and included those in the newly created set of field-cited articles, from which we then sampled negative candidates.
As in dataset-level evaluation, we repeat the candidate sampling procedure three times for all the fields where the field-cited article set is larger than 100k (all except Art, History, and Philosophy) and report the mean values of the metrics.

\begin{table*}[t]
\footnotesize
\setlength{\tabcolsep}{2pt}
\centering
\begin{tabular}{@{}lrrrrrrrrrrrrrrrrrrr|r@{}}
\toprule
& Art & Bio & Bus & Ch & CS & Eco & Eng & ES & Geog & Geol & His & MS & Mat & Med & Phi & Phy & PS & Psy & Soc & AVG \\ \midrule
BM25 & \textbf{36.1} & 44.3 & \textbf{21.2} & \textbf{42.6} & 35.6 & \textbf{26.4} & \textbf{34.2} & \textbf{27.5} & \textbf{29.2} & \textbf{32.1} & \textbf{32.8} & \textbf{34.9} & \textbf{35.6} & 46.3 & \textbf{25.0} & \textbf{35.8} & \textbf{21.2} & 31.9 & \textbf{17.7} & \textbf{32.1} \\
\scibert & 6.2 & 6.7 & 3.2 & 6.7 & 4.2 & 4.7 & 4.6 & 4.8 & 4.8 & 4.7 & 5.3 & 4.4 & 4.9 & 5.4 & 3.5 & 5.5 & 2.8 & 5.0 & 3.0 & 4.8 \\
\specter & 25.1 & 37.0 & 18.0 & 35.4 & 33.7 & 22.6 & 28.5 & 22.6 & 20.1 & 20.3 & 17.4 & 29.0 & 31.5 & 48.6 & 16.2 & 27.5 & 14.4 & 32.0 & 14.3 & 26.0 \\
\scincl & 26.9 & 42.2 & 18.7 & 39.6 & \textbf{37.8} & 23.4 & 31.5 & 25.5 & 23.7 & 22.7 & 20.7 & 30.9 & 33.5 & \textbf{52.4} & 18.7 & 31.4 & 16.5 & \textbf{34.1} & 15.7 & 28.7 \\
\textsc{Aspire-BM} & 26.8 & \textbf{44.7} & 19.3 & 39.9 & 35.3 & 24.3 & 29.7 & 24.5 & 23.3 & 22.9 & 20.1 & 29.8 & 33.1 & 52.1 & 17.2 & 30.0 & 15.9 & 33.1 & 14.3 & 28.2 \\ 
\textsc{Aspire-CS} & 25.8 & 37.1 & 20.0 & 34.9 & 35.8 & 23.5 & 30.2 & 21.9 & 21.7 & 20.1 & 18.4 & 27.5 & 34.2 & 46.5 & 17.1 & 29.0 & 15.3 & 32.7 & 15.7 & 26.7 \\\bottomrule
\end{tabular}
\caption{Results in terms of MAP in the ``field-level'' evaluation setup. Values in \textbf{bold} indicate the best performing model per field. Table with field-abbreviation mapping is given in Table \ref{tab:abbr} in Appendix \ref{sec:appendix}.}
\label{tab:results-field}
\end{table*}

Field-level results in terms of \map are shown in Table~\ref{tab:results-field} (the \ndcg and R@30 results are included in Appendix \ref{sec:appendix}; the best performing models are the same in all cases except R@30 for the Bio field).
As with dataset-level evaluation, we retrieve the top 500 candidates and report results on these sets. In this setup, we also include \textsc{Aspire-CS}, i.e., \aspire variant trained on computer science articles.

BM25 achieves the highest mean \map across all fields, again demonstrating the robust performance of lexical retrieval.
\scincl performs close to BM25, performing best on CS, Med, and Psy fields.
While \scincl's strong performance in CS and Med fields could be explained by a high percentage of \scincl training queries from those fields ($\sim$16\% and $\sim$25.3\% of \scincl train queries come from Med and CS fields, respectively), a high \map value in Psy field is unexpected given the small percentage of Psy queries in \scincl's train set ($\sim$4.1\%).
Analyzing performance across fields, models perform quite well in some fields (e.g., Med and Bio) and worse in others (e.g., Soc, Bus, PS).
Regarding these differences, we note that training sets of most of the TAEs (\specter, \scincl, \textsc{Aspire-BM}) have a highly skewed distribution toward Med, Bio, and CS fields.
However, another possible explanation might be the different levels of interdisciplinarity in particular fields, which could lead to a richer vocabulary than in mono-disciplinary fields.
We leave the investigation of the performance across fields for future work.

Comparing TAEs between each other supports our dataset-level results: \scincl performs slightly better than \aspire (BM), but both outperform \specter, which in turn surpasses \scibert.
Just as in dataset-level evaluation, this ordering is expected given the differences in the training objectives and the training signal used.
When comparing different TAEs across fields, we observe that \aspire performs especially well
in the fields on which it was originally trained: \textsc{Aspire-BM} outperformed other TAEs in Bio and Ch fields, which shows that field-specific sentence-level encoders might be more successful than other TAEs for other fields as well.
Field-level evaluation results also confirm the need for large-scale evaluation of TAEs -- their performance is again much worse than in small-scale benchmark evaluation scenarios, such as \scidocs.

To sum up, both of our setups demonstrated (1) a strong performance of a lexical retrieval model BM25, which either surpassed (field-level) or performed competitively to TAEs (dataset-level) in large-scale evaluation scenarios, and (2) a large decrease in performance of all the evaluated TAEs compared to previous small-scale benchmark setups (\scidocs).
Although we argue that large-scale evaluation is mandatory for more realistic performance estimates, we also recognize the benefits of standardized evaluation benchmarks as they enable the research community to track the improvement on a task easily. 
However, even when evaluation is not performed on a large scale, we argue that to keep the benchmark-obtained performance estimation as realistic as possible, small benchmarks should contain realistic candidate pools with challenging negatives. 
With this in mind, in the next section, we describe the construction of a small but more realistic benchmark for evaluating article representations.

%% file: 4-MultiDomainBenchmark.tex
\section{Multi-Domain Citation Recommendation Benchmark}
\label{sec:md-bench}

We now present our newly constructed \textbf{M}ulti-\textbf{D}omain \textbf{C}itation \textbf{R}ecommendation benchmark -- \textbf{\mdcr}.
As queries in \mdcr, we use the same 200 queries per field as in the field-level evaluation setup (\cref{subsec:field-level-eval}). 
For the candidate pools, we start with a random sampling of 5 articles cited in the query article and then select negative candidates.

\subsection{Benchmark Construction}
\label{subsec:cand_types}
To construct challenging candidate pools, we used four different candidate selection strategies: (1) model-based, (2) graph neighbors-based, (3) citation count-based, and (4) random selection.
Each candidate strategy produces different candidate types that can be used for a more detailed evaluation of the model's performance.
We outline the selection strategies below.

\paragraph{Model-based selection.}
This strategy aims to capture difficult candidates for the models evaluated in the large-scale setup.
As these candidates are difficult for current models, we expect at least some of these candidates to be challenging for some of the future models.
\citet{brown2019large} used a similar method for candidate pool construction in 
\relish, where candidates were selected using three different retrieval models and then annotated for relevance by the field experts.
In \mdcr, we do not provide expert-level annotations for candidates but instead, rely on citations as proxy signals for relevance.

We started with compiling lists of the top 200 candidates per query obtained with each model on the candidate pool from the field-level setup.
As we evaluate models that are both trained and used differently, it is reasonable to expect each model to have difficulties with different negative candidates.
With this in mind, we intended to select those candidates that are difficult for different models.
To determine the degree to which the models' top candidates overlap, we calculated the average Jaccard index between the highest-ranked negative candidates of different model pairs.
The models that obtained a low average Jaccard index tend to make different mistakes (i.e., rank different negative candidates highly) than other models.
We chose the three models with the lowest average Jaccard index for selecting negative candidates in this strategy: BM25, \scincl, and \specter.
For each query and each of the three selected models, we randomly sampled ten negative candidates from the top 200 highest ranked candidates by the model and added them to the query's candidate pool.
We call these candidate types \texttt{BM25}, \texttt{SPECTER}, and \texttt{SciNCL} for candidates obtained from the respective models.

\paragraph{Graph neighbors-based selection.}
Research on citation-seeking behavior states that scientists often traverse citation graphs to find articles relevant to their needs \citep{belter2016citation,hinde2015bidirectional}.
This suggests that challenging articles should be sampled from the same source, i.e., from a set of articles that either cite or are cited in the articles relevant to the query.

To include such candidates in our pools, we employed the following procedure over the citation graph.
Let $q$ be a query article and $\mathit{OC_q} = \{c_1, ..., c_n\}$ a set of articles that are cited in $q$ (i.e., \textit{outgoing} citations). 
For each $c_i \in \mathit{OC_q}$, we constructed corresponding $\mathit{OC_{c_i}}$ and $\mathit{IC_{c_i}} = \{i_1, ..., i_m\}$ sets, where $i_j$ represents an article that cites $c_i$ (i.e., \textit{incoming} citations).
Using such sets, we calculated the overlap similarity as $O_{q,c_i} = | \mathit{OC_q} \cap (\mathit{OC_{c_i}} \cup \mathit{IC_{c_i}}) | / |\mathit{OC_q}|$, which represents the similarity between $q$'s outgoing citations and $c_i$'s incoming and outgoing citations.
The high $O_{q,c_i}$ value suggests a considerable overlap in citation links between $q$ and $c_i$, which indicates that these articles are highly topically related.

We calculated $O_{q,c_i}$ for all the query articles and their cited articles.
We then sorted the cited articles by their $O_{q,c_i}$ values, starting from the highest (highly topically relevant) to the lowest (slightly topically related).
Since we wanted to make our candidate pool challenging, we started with the $c_i$ that has the highest $O_{q,c_i}$ value and added to the query's candidate pool all the articles from its $\mathit{OC_{c_i}} \cup \mathit{IC_{c_i}}$ set that are not in $\mathit{OC_q}$ (i.e., cited in the query article).
We repeated this procedure until ten negative candidates were added to the pool.
We call this candidate type \texttt{Graph}.

\paragraph{Citation count-based selection.}
In this selection strategy, we created a list of the top 200 most cited articles in each scientific field.
We used \storc's MAG field to detect articles from each field and sorted them by the citation counts in descending order.
We then randomly sampled ten candidate articles for each query article based on the query article's MAG field and added these articles to the candidate pool.
This type of candidates is called \texttt{Most} \texttt{cited}.

\paragraph{Random selection.}
Finally, as a less challenging and baseline candidate set, we settled for a random selection strategy, where we randomly sampled ten candidates from the prefiltered set of \storc articles.
We call this candidate type \texttt{Random}.

\subsection{Benchmark Size}
Overall, \mdcr contains 200 queries per each of the 19 MAG fields, where each query is paired with a set of 60 negative candidates and five cited articles, totalling 247,000 query-candidate pairs that need to be evaluated. 
Compared to \scidocs, where 1,000 queries are paired with candidate pools of size 25 (a total of 25,000 query-candidate pairs), \mdcr is almost ten times bigger.
While this growth in size increases the computational complexity when using \mdcr compared to other smaller benchmarks, it arguably makes the results more realistic. 
In addition, we also note that since \mdcr is split across different scientific fields, models can be evaluated on specific fields only, which reduces the number of query-candidate pairs to be evaluated.

\subsection{Results}

\begin{table*}[h]
\footnotesize
\setlength{\tabcolsep}{7.35pt}
\centering
\begin{tabular}{@{}lrr|rr|rr|rr|rr|rr@{}}
\toprule
Models $\rightarrow$ & \multicolumn{2}{c}{BM25} & \multicolumn{2}{c}{\scibert} & \multicolumn{2}{c}{\specter} & \multicolumn{2}{c}{\scincl} & \multicolumn{2}{c}{\textsc{Aspire-BM}} & \multicolumn{2}{c}{\textsc{Aspire-CS}} \\ 
\cmidrule(lr){2-3}\cmidrule(lr){4-5}\cmidrule(lr){6-7}\cmidrule(lr){8-9}\cmidrule(lr){10-11}\cmidrule(lr){12-13}
Fields $\downarrow$ & \map & R@5 & \map & R@5 & \map & R@5 & \map & R@5 & \map & R@5 & \map & R@5 \\
\midrule
Art & \textbf{38.2}	&	\textbf{32.3}	&	22.4	&	16.6	&	34.1	&	28.8	&	34.7	&	29.2	&	34.0	&	27.7	&	34.1	&	28.0 \\ 
Bio & 38.3	&	33.6	&	20.4	&	14.0	&	34.6	&	30.0	&	36.8	&	32.3	&	\textbf{38.7}	&	\textbf{33.7}	&	35.7	&	29.9 \\
Bus & 28.1	&	22.5	&	19.1	&	13.1	&	27.5	&	21.8	&	28.5	&	\textbf{24.6}	&	28.5	&	23.4	&	\textbf{29.6}	&	23.1 \\
Ch & \textbf{38.0}	&	\textbf{32.6}	&	20.0	&	13.7	&	33.7	&	29.3	&	36.5	&	31.5	&	36.5	&	31.0	&	34.1	&	28.3 \\
CS & 34.8	&	30.5	&	19.5	&	12.7	&	35.6	&	30.4	&	\textbf{37.2}	&	\textbf{32.2}	&	35.4	&	30.4	&	35.4	&	30.1 \\
Eco & \textbf{30.5}	&	\textbf{26.0}	&	21.4	&	15.4	&	27.3	&	21.9	&	28.3	&	23.2	&	29.3	&	24.3	&	28.0	&	22.7 \\
Eng & \textbf{34.6}	&	\textbf{29.3}	&	20.5	&	13.9	&	31.3	&	27.3	&	34.2	&	28.0	&	32.7	&	27.7	&	33.4	&	28.1 \\
ES & \textbf{31.6}	&	\textbf{26.2}	&	21.3	&	15.1	&	30.1	&	24.2	&	31.5	&	25.5	&	30.8	&	24.7	&	29.9	&	23.7 \\
Geog & \textbf{31.8}	&	\textbf{27.8}	&	21.9	&	16.7	&	26.4	&	22.2	&	29.5	&	23.8	&	30.3	&	26.0	&	28.4	&	22.2 \\
Geol & \textbf{33.1}	&	\textbf{28.0}	&	19.5	&	13.9	&	24.8	&	20.1	&	25.7	&	19.9	&	28.5	&	23.5	&	25.8	&	21.4 \\
His & \textbf{38.1}	&	\textbf{32.9}	&	20.8	&	15.2	&	27.1	&	20.6	&	30.9	&	23.9	&	31.0	&	24.2	&	28.5	&	22.1 \\ 
MS & \textbf{36.1}	&	\textbf{30.7}	&	22.1	&	15.5	&	34.1	&	28.2	&	35.8	&	29.6	&	35.8	&	29.8	&	34.0	&	29.2 \\
Mat & 35.3	&	28.3	&	22.8	&	18.3	&	34.2	&	28.9	&	34.9	&	30.1	&	36.2	&	31.0	&	\textbf{36.9}	&	\textbf{32.2} \\
Med & 38.6	&	32.5	&	22.0	&	16.4	&	41.4	&	36.3	&	42.7	&	36.5	&	\textbf{44.0}	&	\textbf{37.8}	&	41.7	&	36.7 \\
Phi & \textbf{30.2}	&	\textbf{25.7}	&	19.2	&	13.3	&	27.1	&	21.1	&	29.9	&	23.5	&	28.7	&	24.1	&	29.1	&	23.3 \\
Phy & \textbf{35.1}	&	30.2	&	23.9	&	18.1	&	30.8	&	26.3	&	34.5	&	\textbf{30.3}	&	32.9	&	27.7	&	32.9	&	28.7 \\
PS & \textbf{28.6}	&	\textbf{23.1}	&	19.4	&	14.0	&	24.2	&	18.0	&	26.4	&	21.7	&	25.9	&	21.2	&	26.8	&	21.7 \\
Psy & 32.5	&	28.9	&	20.3	&	16.2	&	32.3	&	28.1	&	34.2	&	\textbf{30.5}	&	\textbf{34.3}	&	29.4	&	34.2	&	28.3 \\
Soc & 26.8	&	20.5	&	20.2	&	15.8	&	25.2	&	20.5	&	26.7	&	21.9	&	\textbf{27.3}	&	\textbf{22.2}	&	26.7	&	\textbf{22.2} \\
\midrule
AVG & \textbf{33.7} & \textbf{28.5} & 20.9 & 15.2 & 30.6 & 25.5 & 32.6 & 27.3 & 32.7 & 27.4 & 31.8 & 26.4 \\
\bottomrule
\end{tabular}
\caption{Results in terms of \map and R@5 on \mdcr. Values in \textbf{bold} indicate the best performing model for a combination of field and metric.}
\label{tab:results-benchmark-v2}
\end{table*}

Results of evaluation on \mdcr are given in Table~\ref{tab:results-benchmark-v2}.
We report \map and R@5 (each query is coupled with five positive candidates) across all pairs of the scientific fields and evaluated the model.
We evaluate the same set of models as in the field-level large-scale evaluation.

Results demonstrate, yet again, a strong performance from BM25, which outperformed all other models in terms of average metric scores across all fields.
Interestingly, when evaluated on \mdcr's small-sized pools, the difference in performance between \scibert and other TAEs (e.g., \scincl) is smaller than in large-scale evaluation (11.7 in \map on \mdcr vs. 37 in \map on dataset-level, 200k pool size).
Such a difference in results confirms the benefits of evaluating TAEs on larger pools to obtain more realistic results.
Another observation is a similar average performance between \scincl and \aspire, despite \aspire variants being trained only on the articles from specific fields (biomedicine and computer science).
As in the field-level evaluation, competitive results from \aspire indicate that sentence-level representations might be able to capture a more informative signal between related articles than document-level ones.

Although BM25 outperforms other models in most fields, TAEs obtain the best scores in some cases when looking at performance in specific fields.
Specifically, \textsc{Aspire-BM} is the top-performing model in the Bio, Med, and Soc fields (and Psy in \map value), which is not surprising as it was trained on articles from the biomedicine field.
Similar goes for \textsc{Aspire-CS} and its performance in the Mat field, although it does not yield the best results in the field it was trained on (CS).
However, when analyzing \aspire's performance, it is worth noting that we did not account for the overlap of \aspire training queries with our new benchmark since \aspire's training set was not publicly available at the time of writing.
For this reason, the results of both \aspire variants might be too optimistic if the train-test overlap is significant.

\subsection{Performance across Candidate Types}

\begin{table*}[h]
\footnotesize
\setlength{\tabcolsep}{5.7pt}
\centering
\begin{tabular}{@{}lrr|rr|rr|rr|rr|rr@{}}
\toprule
Candidate types $\rightarrow$ & \multicolumn{2}{c}{\texttt{BM25}} & \multicolumn{2}{c}{\texttt{SPECTER}} & \multicolumn{2}{c}{\texttt{SciNCL}} & \multicolumn{2}{c}{\texttt{Graph}} & \multicolumn{2}{c}{\texttt{Most cited}} & \multicolumn{2}{c}{\texttt{Random}} \\ 
\cmidrule(lr){2-3}\cmidrule(lr){4-5}\cmidrule(lr){6-7}\cmidrule(lr){8-9}\cmidrule(lr){10-11}\cmidrule(lr){12-13}
Models $\downarrow$ & \map & R@5 & \map & R@5 & \map & R@5 & \map & R@5 & \map & R@5 & \map & R@5 \\
\midrule
BM25 & 52.2 & 39.0 & 68.8 & 57.5 & 68.0 & 56.9 & 58.0 & 46.7 & 90.3 & 82.2 & 93.0 & 86.3 \\
\scibert & 50.5 & 39.1 & 47.6 & 36.0 & 49.0 & 37.7 & 47.2 & 36.1 & 79.6 & 69.3 & 84.9 & 75.8 \\
\specter & 67.4 & 56.3 & 51.1 & 38.7 & 57.8 & 45.8 & 57.3 & 46.3 & 92.8 & 86.5 & 99.0 & 97.1 \\
\scincl & 68.3 & 57.6 & 61.2 & 49.8 & 54.1 & 42.1 & 58.1 & 47.3 & 94.0 & 88.3 & 99.0 & 97.2 \\
\textsc{Aspire-BM} & 66.7 & 55.6 & 57.7 & 46.5 & 59.3 & 47.5 & 57.3 & 46.5 & 93.6 & 87.6 & 99.1 & 97.2 \\
\textsc{Aspire-CS} & 66.0 & 54.9 & 55.7 & 43.8 & 58.5 & 46.8 & 57.2 & 46.6 & 94.3 & 88.7 & 98.9 & 96.8 \\
\midrule
AVG & 61.8 & 50.4 & 57.0 & 45.4 & 57.8 & 46.1 & 55.9 & 44.9 & 90.8 & 83.8 & 95.7 & 91.7 \\
\bottomrule
\end{tabular}
\caption{Results in terms of \map and R@5 for different candidate types on \mdcr.}
\label{tab:results-cands}
\end{table*}

To analyze the difficulty of candidate types that we introduced in \cref{subsec:cand_types}, we evaluate the models on subsets of candidate pools consisting of 5 cited articles and all negative candidates from specific candidate type.
Evaluation across candidate types allows us to analyze how difficult each candidate type is for each model. 
As the candidates obtained via model-based selection are chosen precisely because they were difficult for the particular models, we do expect these models to not perform well on such candidates. However, such evaluation can reveal interesting insights into the differences across the evaluated models, e.g., whether the same candidate types are difficult for all neural-based models.

Results of this evaluation are presented in Table~\ref{tab:results-cands}.
Unsurprisingly, \texttt{Random} candidates are the easiest candidate type for all the evaluated models.
Candidates from the \texttt{Most} \texttt{cited} type are also relatively easy for the models, with on average $>$90 score in \map.
On average, the most challenging candidate type is the \texttt{Graph} candidates subset, with an average \map score of 55.9.
Interestingly, the best-performing model on the \texttt{Graph} candidates subset is \scincl, which explicitly uses citation graph embeddings in selecting training examples.
Such a training strategy seems to help the model distinguish between relevant and irrelevant graph neighbors.

The performance on the candidate types obtained with the model-based selection strategy differs between TAEs and BM25, which is somewhat expected given the difference between neural (TAEs) and non-neural (BM25) models.
As expected, negative candidates from the \texttt{BM25} type are the most difficult for BM25 itself since those were sampled from a set of top candidates provided by BM25. 
On the other hand, TAEs (\scibert excluded) all perform similarly well on the \texttt{BM25} candidate type.
Likewise, \texttt{SPECTER} and \texttt{SciNCL} candidates are the most difficult for \specter and \scincl, respectively, while BM25 performs better than TAEs on these candidate types. 
It is interesting to note the difference in the performance of \scincl on \texttt{SPECTER} candidates compared to the performance of \specter on \texttt{SciNCL} candidates. 
While \scincl outperforms \specter on \texttt{SPECTER} candidates with more the 10 points in the absolute value (for both metrics), \specter improves over \scincl on \texttt{SciNCL} candidates with only 3.7 absolute points.
These results again confirm that the way in which negative candidates are sampled when training the models with the contrastive learning objective is important.
As for the \aspire variants and candidate types obtained with the model-based strategy, \aspire variants perform better on the \texttt{BM25} candidate type than on the \texttt{SPECTER} or \texttt{SciNCL} type.
We hypothesize that such difference is due to TAEs being similar neural models and therefore prone to similar errors regarding semantic vs.~lexical matching of texts. 
In contrast, BM25, a purely lexical model, makes different errors.
We leave the analysis of the differences in performance between neural and non-neural models for future work.

%% file: 5-Conclusion.tex
\section{Conclusion}
\label{sec:concl}

We evaluated transformer-based article encoders in large-scale citation recommendation scenarios %retrieval scenarios \jan{retrieval $\to$ scientific article recommendation? Or, better, talk about citation recommendation? At any rate, make sure to align with the title, abstract, and intro.} 
across different scientific fields and candidate pool sizes.
Together with transformer-based encoders, we evaluated the performance of a robust lexical retrieval baseline BM25 and demonstrated that it still performs competitively with recent neural-based models.
In the case of large field-specific candidate pools, BM25 outperformed transformer-based models in most fields.

Furthermore, to promote a more realistic and a more diverse evaluation across different fields in comparison to the existing benchmarks used for evaluating scientific article representations, we presented a new multi-domain benchmark dataset based on citation recommendation task, which we call \mdcr. 
Evaluation on \mdcr demonstrated the difficulty of specific candidate types and set the ground for evaluating future scientific article encoders.

Our evaluation demonstrated the varying performance across scientific fields, which we believe should be analyzed in future work to improve encoders' performance across all fields, not just those prevailing in the datasets.
Given that our benchmark dataset is not expert-annotated but rather based on citations as relevance signals, we propose constructing an expert-annotated dataset with articles from different scientific fields.
We hope our contributions will stimulate the community to work on more realistic and challenging evaluation setups of scientific article recommendation models. %\jan{for scientific article recommendation}?

%% file: 6-Appendix.tex
\clearpage

\appendix

\section{Appendix}
\label{sec:appendix}

% \jan{vidi možeš li ukrotiti appendix da bude na jednoj stranici, ili na više, ali bez tolikih praznina između tablica. Mo\eš forsirati pozicije tablica sa h!}

\subsection{BM25}

Here we provide details about the relevance score function used by BM25 \citep{robertson1994some}.
Before calculating relevance scores for pairs of articles, article texts are first transformed into bag-of-words vectors. 
Given a query $Q$, containing terms $q_1, ..., q_n$, and a document $D$, BM25 calculates relevance score $\mathrm{s}$ as follows:
\begin{multline*} 
%\label{eq:bm25}
\small
\mathrm{s}(Q,D) = \\ \sum_{i=1}^{n} \mathrm{IDF}(q_i) \cdot \frac{f(q_i,D)\cdot(k_1 + 1)}{f(q_i,D) + k_1 \cdot \left(1 - b + b \cdot \frac{|D|}{\mathrm{avgdl}}\right)}
\end{multline*}
where $f(q_i,D)$ is the frequency $q_i$ in document $D$, $|D|$ is the length of $D$ in words, $\mathrm{avgdl}$ is the average document length, and $k_1$ and $b$ are parameters that can be tuned for a specific document collection. 
$\mathrm{IDF}(q_i)$ is the inverse document frequency for $q_i$, and is typically calculated as:
\begin{equation*}
\small
\mathrm{IDF}(q_i) = \ln\left(\frac{N - n(q_i) + 0.5}{n(q_i) + 0.5} + 1\right)
\end{equation*}
where $N$ is the total number of documents in the collection  and $n(q_i)$ is the number of documents containing the term $q_i$. 

Intuitively, $\mathrm{s}$ will output higher scores for a document $D$ that contains many terms as $Q$, that also do not appear often in other documents.
When it comes to parameters $b$ and $k_1$, $b$ controls to what degree the length of a document will affect the final score \citep{lipani2015verboseness}, while $k_1$ controls to what degree an additional occurrence of a term affects the final score \citep{lv2011adaptive}.

\subsection{Field Abbreviations}

\begin{table}[h!]
\footnotesize
\centering
\begin{tabular}{@{}r|l@{}}
\toprule
MAG field & Abbreviation \\ \midrule
Art & Art \\
Biology & Bio \\ 
Business & Bus \\
Chemistry & Ch \\
Computer Science & CS \\
Economics & Eco \\
Engineering & Eng \\
Environmental Science & ES \\
Geography & Geog \\
Geology & Geol \\
History & His \\
Materials Science & MS \\
Mathematics & Mat \\
Medicine & Med \\
Philosophy & Phi \\
Physics & Phy \\
Political Science & PS \\
Psychology & Psy \\
Sociology & Soc \\
\bottomrule
\end{tabular}
\caption{Abbreviations for MAG fields that we use in the field-level evaluation and in the new benchmark.}
\label{tab:abbr}
\end{table}

Table~\ref{tab:abbr} shows the abbreviations for MAG fields.

\subsection{Evaluation on \scidocs}

\begin{table}[h!]
\footnotesize
\setlength{\tabcolsep}{7pt}
\centering
\begin{tabular}{@{}rrr@{}}
\toprule
Model & \map & \ndcg \\ \midrule
\scibert & 48.3 & 71.7 \\
\specter & 88.3 & 94.9 \\
\scincl & 93.6 & 97.3 \\
\textsc{ts-Aspire} & 91.0 & 95.0 \\
\bottomrule
\end{tabular}
\caption{Results of different TAEs evaluated on \scidocs's ``Cite'' task. Values as reported in \citep{cohan-etal-2020-specter}, \citep{mysore2021aspire}, and \citep{ostendorff2022neighborhood}.}
\label{tab:results-scidocs}
\end{table}

Table~\ref{tab:results-scidocs} shows the results of evaluation of \scibert, \specter, \scincl, and \aspire on \scidocs benchmark, as reported in the previous work.

\subsection{Field-level Evaluation Results}

\begin{table*}[t]
\footnotesize
\setlength{\tabcolsep}{2pt}
\centering
\begin{tabular}{@{}r|rrrrrrrrrrrrrrrrrrr|r@{}}
\toprule
Model & Art & Bio & Bus & Ch & CS & Eco & Eng & ES & Geog & Geol & His & MS & Mat & Med & Phi & Phy & PS & Psy & Soc & AVG \\ \midrule
BM25 & \textbf{64.2} & 78.8 & \textbf{56.9} & \textbf{76.2} & 69.3 & \textbf{65.0} & \textbf{68.0} & \textbf{62.6} & \textbf{64.5} & \textbf{69.2} & \textbf{62.7} & \textbf{69.8} & \textbf{70.1} & 77.8 & \textbf{56.9} & \textbf{70.1} & \textbf{56.2} & 70.2 & \textbf{54.0} & \textbf{66.5} \\
\scibert & 33.2 & 45.8 & 33.6 & 43.7 & 36.0 & 39.6 & 36.4 & 36.8 & 37.6 & 39.6 & 32.9 & 37.3 & 37.9 & 39.8 & 31.2 & 39.5 & 32.1 & 40.9 & 32.9 & 37.2 \\
\specter & 53.5 & 74.2 & 53.9 & 71.8 & 67.7 & 61.9 & 63.6 & 58.2 & 56.5 & 59.2 & 47.0 & 65.6 & 66.5 & 79.2 & 48.6 & 63.8 & 49.2 & 70.5 & 49.9 & 61.1 \\
\scincl & 55.3 & 77.5 & 54.1 & 74.0 & \textbf{70.3} & 62.5 & 66.0 & 60.7 & 59.6 & 61.2 & 50.5 & 67.0 & 68.0 & \textbf{81.1} & 50.8  & 67.1 & 51.4 & \textbf{71.7} & 51.3 & 63.1 \\
\textsc{Aspire-BM} & 55.6 & \textbf{79.2} & 55.2 & 74.4 & 69.1 & 63.5 & 64.8 & 60.0 & 59.4 & 61.5 & 50.1 & 66.2 & 68.1 & 81.0 & 50.1 & 66.2 & 50.6 & 70.9 & 49.8 & 62.9 \\ 
\textsc{Aspire-CS} & 54.8 & 74.2 & 56.1 & 71.1 & 69.4 & 62.9 & 65.0 & 57.7 & 57.8 & 59.0 & 48.1 & 64.3 & 69.1 & 77.9 & 49.6 & 65.5 & 50.4 & 70.8 & 51.4 & 61.8 \\\bottomrule
\end{tabular}
\caption{Results in terms of \ndcg in the ``field-level'' evaluation setup. Values in \textbf{bold} indicate the best performing model per field.}
\label{tab:results-field-ndcg}
\end{table*}

\begin{table*}[t]
\footnotesize
\setlength{\tabcolsep}{2pt}
\centering
\begin{tabular}{@{}r|rrrrrrrrrrrrrrrrrrr|r@{}}
\toprule
Model & Art & Bio & Bus & Ch & CS & Eco & Eng & ES & Geog & Geol & His & MS & Mat & Med & Phi & Phy & PS & Psy & Soc & AVG \\ \midrule
BM25 & \textbf{46.5} & \textbf{38.9} & \textbf{27.5} & \textbf{41.7} & 43.1 & \textbf{29.7} & \textbf{42.7} & \textbf{35.3} & \textbf{34.6} & \textbf{34.2} & \textbf{42.7} & \textbf{39.5} & \textbf{42.2} & 48.2 & \textbf{34.0} & \textbf{40.9} & \textbf{28.3} & 32.0 & \textbf{24.0} & \textbf{37.2} \\
\scibert & 9.0 & 4.8 & 2.1 & 5.8 & 5.0 & 3.3 & 5.5 & 5.1 & 5.0 & 3.9 & 8.0 & 4.1 & 6.0 & 4.8 & 3.8 & 6.6 & 1.6 & 3.4 & 2.4 & 4.8 \\
\specter & 35.5 & 33.4 & 24.7 & 35.3 & 41.9 & 25.7 & 36.4 & 30.3 & 24.3 & 23.4 & 24.8 & 34.2 & 38.5 & 49.0 & 24.1 & 33.4 & 19.9 & 32.6 & 19.8 & 30.9 \\
\scincl & 38.9 & 36.8 & 25.4 & 39.3 & \textbf{46.4} & 26.0 & 39.5 & 33.0 & 28.0 & 25.5 & 28.2 & 35.5 & 40.2 & \textbf{52.0} & 26.4 & 37.1 & 23.0 & \textbf{34.7} & 21.5 & 33.5 \\
\textsc{Aspire-BM} & 34.9 & 38.2 & 25.1 & 38.6 & 42.5 & 27.3 & 37.0 & 31.5 & 27.5 & 25.3 & 28.0 & 34.5 & 39.9 & 51.6 & 25.0 & 35.2 & 22.2 & 32.3 & 19.2 & 32.4 \\ 
\textsc{Aspire-CS} & 34.9 & 33.0 & 25.5 & 34.7 & 42.1 & 25.9 & 37.5 & 29.4 & 25.6 & 23.2 & 25.3 & 33.0 & 40.1 & 47.3 & 24.7 & 34.5 & 21.4 & 32.2 & 20.1 & 31.1 \\\bottomrule
\end{tabular}
\caption{Results in terms of recall@30 in the ``field-level'' evaluation setup. Values in \textbf{bold} indicate the best performing model per field.}
\label{tab:results-field-r30}
\end{table*}

Results for ``field-level'' evaluation setup in terms of \ndcg and recall@30 are given in Tables \ref{tab:results-field-ndcg} and \ref{tab:results-field-r30}, respectively.
Best scoring combinations of model and field are mostly the same as in case of \map (reported in Table \ref{tab:results-field}), with the exception of recall@30 in Bio field, where BM25 yields the best result (as opposed to \textsc{Aspire-BM} in case of \map).